\documentclass[preprint,12pt,authoryear]{elsarticle}

\usepackage{titlesec}
\titleformat{\paragraph}
{}{\theparagraph}{1em}{}
\titlespacing*{\paragraph}
{0pt}{3.25ex plus 1ex minus .2ex}{1.5ex plus .2ex}

\usepackage{amssymb}
\usepackage{amsmath}

\usepackage{rotating}

\journal{Nuclear Physics B}

\begin{document}

\begin{frontmatter}



\title{Measurement of Trace Elements in Volcanic Materials: Consequences for the Cretaceous-Tertiary Mass Extinction, Geoneutrinos and the Origin of the Hawaii's Archipelago} 


\author[inst1]{Pedro V. Guillaumon\corref{cor1}}
\author[inst1]{Iuda D. Goldman\fnref{label1}}
\author[inst2,inst3]{Eric B. Norman\corref{cor2}}
\author[inst2,inst3,inst5]{Keenan J. Thomas}
\author[inst1]{Paulo R. Pascholati}
\author[inst2]{Ross E. Meyer}
\author[inst2]{Jordan L. Sabella}
\author[inst2]{Alan R. Smith\fnref{label1}}

\affiliation[inst1]{organization={Universidade de Sao Paulo,Instituto de Fisica},
            addressline={Rua do Matao, 1371}, 
            city={Sao Paulo},
            postcode={05508090}, 
            state={Sao Paulo},
            country={Brazil}}

\affiliation[inst2]{organization={Department of Nuclear Engineering, University of California, Berkeley},
            addressline={4153 Etcheverry Hall}, 
            city={Berkeley},
            postcode={94720}, 
            state={California},
            country={USA}}

\affiliation[inst3]{organization={Lawrence Berkeley National Laboratory},
            addressline={1 Cyclotron Rd}, 
            city={Berkeley},
            postcode={94720}, 
            state={California},
            country={USA}}

\affiliation[inst5]{organization={Current address: Lawrence Livermore National Laboratory},
            addressline={7000 East Ave}, 
            city={Livermore},
            postcode={94550}, 
            state={California},
            country={USA}}

\cortext[cor1]{Corresponding author: guillaumon@if.usp.br}
\cortext[cor2]{Corresponding author: ebnorman@lbl.gov}
\cortext[cor3]{$^1$Deceased}

\begin{abstract}
Seventeen representative samples of volcanic origin were collected from Ecuador (Pichincha Volcano), Iceland  (Eyjafjallajökull Volcano), India (Deccan Traps), Hawaii, Kilimanjaro, Mt. Etna, Rwanda (Virunga Mountains), and Uganda (Virunga Mountains). Neutron activation analysis (NAA) was performed to determine the concentration of 33 chemical elements, including 21 trace elements, 20 heavy metals, and 9 rare earth elements: Al, As, Ba, Ca, Ce, Cl, Co, Cr, Cs, Dy, Eu, Fe, Hf, K, La, Lu, Mg, Mn, Na, Nd, Rb, Sb, Sc, Sm, Sr, Ta, Tb, Th, Ti, U, Yb, Zn, and Zr.

Correlation analysis of the abundance of samples from different islands in the Hawaii archipelago (Kauai, Kilauea, Mauna Loa, and Haleakala) confirmed that the islands were likely formed by two different lava sources. Additionally, the upper limit of iridium was determined in 11 of these samples using Bayesian analysis, which does not support the hypothesis that volcanic activity caused the extinction of the dinosaurs.

We also discuss how the abundance of thorium and uranium in  lava from different geological formations and depths can contribute to building a better map of natural radioisotope occurrences on Earth, which is important for geoneutrino experiments. A high abundance of rare elements was reported in some of the analyzed locations, indicating potential commercial interest and the possibility of exploring volcanoes as sources of chemical elements used in electronic devices.
\end{abstract}

\begin{graphicalabstract}
\end{graphicalabstract}

\begin{highlights}
\item Geoneutrinos
\item Volcanic Analysis
\item Hawaii's origin
\end{highlights}


\end{frontmatter}



\section{Introduction}
\label{sec:intro}

The study of the abundance of chemical elements in lava materials provides crucial information about the structure of the Earth's mantle. Historically, the analysis of chondritic meteorites and other extraterrestrial materials, along with seismic wave and Earth's magnetic field analysis, has been used to infer the Earth's inner structure \citep{Drake2002}. However, it is well known that the mantle's composition differs from these materials, and seismic waves only provide a broad view of the Earth's core components. Therefore, a systematic study of the main and trace elements in materials expelled by different volcanoes worldwide at various epochs is essential for a more detailed "radiography" of the Earth. By analyzing the anisotropies in these samples, it is possible to gain insights into the dynamics of magma. Fluid dynamic models, electrical conductivity, and thermal expansivity measurements indicate that the upper and lower mantle have similar compositions, especially for Mg, Si, Fe, and C composites \citep{Drake2002,Wood_1991,Shim2001,Chopelas_1992,Kellogg1999,Tackley_1993}, making the study of the upper mantle a valuable tool for understanding the Earth's depths. Additionally, this can help understand the role of mantle plumes in forming the Pacific hotspots, whose activity differs significantly from other lava sources \citep{Koppers2021,DeFelice2019,Jones2017}.

Among the possible studies related to chemical anisotropies is the origin of Hawaii. In this archipelago, the volcanoes in the north and south exhibit distinct geochemical behaviors attributed to different lava sources \citep{Huang_2011}. Measurements of ${}^{208}Pb/{}^{206}Pb$, ${}^{87}Sr/{}^{86}Sr$, ${}^{143}Nd/{}^{144}Nd$, and ${}^{176}Hf/{}^{177}Hf$ asymmetries are the main indicators of these different origins \citep{Huang_2011,Abouchami2005,Huang2005,Gao2022,Xu2014,Jones2016}. However, there is still a lack of information about the abundance of elements in the Pacific hotspots that can provide experimental data about them.

This paper also addresses the hypothesis of dinosaur extinction due to volcanic activity. The most accepted theory, proposed by Alvarez et al. \citep{Alvarez}, suggests that the Cretaceous-Tertiary mass extinction was caused by a meteorite impact, as evidenced by the Chicxulub crater. The amount of iridium found there is only compatible with meteorites, as this element is highly depleted on Earth \citep{Hatsukawa2011,Zajzon2011}. Alternative hypotheses suggest that volcanic activity, such as the Deccan volcanism, was the main cause of this mass extinction \citep{Keller2011,Glasby1996,Brett1992}. The dust from the meteorite impact caused the earth to cool, killing the plants and a series of successive events that lead to dinosaurs extinction. This impact could have triggered volcanic activity, although this would be a secondary effect of the Cretaceous-Tertiary mass extinction. One possible piece of evidence against this is if volcanoes can expel an amount of iridium compatible with meteorites. Thus, measuring trace elements, particularly iridium, in lava materials could provide important constraints on volcanic activities during this period.

Furthermore, one of the biggest questions in geology is the Earth's internal heat budget. The internal heat has two main components: radiogenic heat from uranium and thorium decay chains and primordial heat. Although this heat is estimated at $47 (2) \, TW$ \citep{Davies2010}, the fraction due to radiogenic or primordial heat is not well known. If all thorium and uranium are assumed to be in the crust or homogeneously distributed on Earth, the $47 \, TW$ cannot be explained \citep{Davies2010,Dye2012}. One promising technique is the recent measurement of geoneutrinos, i.e., neutrinos emitted in the uranium and thorium decay chains \citep{Agostini}. Since neutrinos are almost transparent to Earth's materials, they could reveal the total amount of uranium and thorium. However, they cannot reveal their concentration locations. Systematic studies of thorium and uranium in the mantle, along with geoneutrino measurements, could potentially solve this question.

Lastly, rare earth elements are essential for electronics  devices like computers. Although required for all modern applications, their rarity contributes to the high prices of these gadgets. Some studies suggest that volcanoes could be a source of rare earth elements \citep{Deady2019}, but only a systematic study of volcanic materials can determine if volcanic exploration is commercially feasible.

This study aimed to systematically measure trace elements in volcanic materials of different origins and locations. By proposing an original correlation analysis, it was possible to study the hypothesis of the Hawaiian archipelago's origin. Additionally, a Bayesian analysis was conducted to estimate the amount of iridium in these samples and investigate the Cretaceous-Tertiary mass extinction due to volcanic activity. Thorium and uranium abundance were measured to provide constraints on the Earth's internal heat components and geoneutrinos. Furthermore, measurements of rare earth element enrichment could make commercial exploration of volcanoes possible. Finally, the measurement of 33 chemical elements, including 21 trace elements in 17 samples, provides important constraints for a better understanding of the Earth's mantle inhomogeneities. Neutron activation analysis (NAA) with gamma spectroscopy was used for these measurements, allowing high accuracy and sensitivity for trace elements.

\section{Materials and Methods}

\subsection{Materials}

Seventeen samples of magmatic origin were collected from different sites: two from Ecuador (Pichincha volcano), two from Iceland, two from India (Deccan Traps), four from Hawaii (Kauai, Kilauea, Mauna Loa, and Haleakala), two from Kilimanjaro, two from Mt. Etna, one from Rwanda (Virunga Mountains), and two from Uganda (Virunga Mountains) (Table~\ref{tab:samples}).

\begin{table}[t]
\centering
\begin{tabular}{l c }
\textbf{Sample Name}	& \textbf{Description} \\
Ecuador I		& From Pichincha volcano 		\\
Ecuador II		& From Pichincha volcano			\\
Iceland I        & Eyjafjallajökull volcano \\
Iceland II      & Eyjafjallajökull volcano \\
India I         & From Deccan Traps \\
India II        & From Deccan Traps \\
Kauai           & From Kauai volcano in Hawaii archipelago \\
Kilauea         & From Kilauea volcano in Hawaii main island \\
Mauna Loa       & From Mauna Loa volcano in Hawaii main island \\
Haleakala       & From Haleakala volcano in Hawaii archipelago (Maui island)  \\
Kilimanjaro I   & From Mt. Kilimanjaro \\
Kilimanjaro II  & From Mt. Kilimanjaro \\
Etna I          & From Mt. Etna \\
Etna II         & From Mt. Etna \\
Rwanda          &  From Virunga Mountains (Rwanda side) \\
Uganda I        & From Virunga Mountains (Uganda side) \\
Uganda II       & From Virunga Mountains (Uganda side) \\
\end{tabular}
\caption{List and description of samples used in this study. \label{tab:samples}}
\end{table}

All the samples collected were at least 15x10 cm to ensure their representativeness. Samples from the same location but from different types of lava were analyzed separately, such as Ecuador I and II. All samples were of magmatic origin and were collected from the surface of the respective volcanoes.

\subsection{Methods}

\subsubsection{Preparation}

\hfill 

The samples were triple cleaned in an ultrasonic cleaner (UC), first with water for 5 minutes, then with a 7\% HF solution and then with water. To obtain a representative sample, all the different crystalline parts of the sample, such as feldspars, were ground together in a mortar. The samples were dried overnight, and then a few grams of the material were encapsulated and sealed for neutron irradiation.

\subsubsection{Neutron Irradiation}

The samples were irradiated with thermal neutrons in the McClellan reactor at the University of California, Davis, and the flux was monitored using a standard pottery as described in Section~\ref{subsec:pottery}.

\hfill 

\paragraph{\textit{Hawaiian and Kilimanjaro samples}}

Two sets of Hawaiian and Kilimanjaro samples were prepared. The first set was irradiated for 10 seconds with a flux of $8.3 (3) \cdot 10^{12} \, n.cm^{-2}.s^{-1}$. The samples were cooled down for 30 minutes and then placed 10 cm from an HPGe detector where gamma spectroscopy was conducted. Several measurements of approximately 10 minutes were performed in order to have a simultaneous measurement of the half-lives fo the isotopes.

A second batch was irradiated for 10 hours with a flux of $4.6 (6) \cdot 10^{11} \, n.cm^{-2}.s^{-1}$. The samples were cooled down for 7 days, and after that, they were placed 12 cm from an HPGe detector, where gamma spectroscopy was performed. The measurements varied from 20 minutes to 2 days over 3 months to follow the half-lives.

\hfill

\paragraph{\textit{Other samples}}

The other samples were irradiated for 8 hours with a flux of $4.7 (5) \cdot 10^{11} \, n.cm^{-2}.s^{-1}$. The samples were cooled down for 7 days and then measured in different HPGe set-ups (simple, in anti-coincidence, or coincidence) with different acquisition times ranging from tens of minutes to days.

\subsubsection{Standard Pottery}
\label{subsec:pottery}

A standard pottery of well-known composition produced by Perlman and Asaro \citep{PERLMAN1969,Smith_1996} was irradiated together with the samples. If a certain isotope is present in both the pottery and the sample, the abundance can be calculated by

\begin{equation}
    \frac{q_{sample}}{q_{sp}} = \frac{C_{sample}}{C_{sp} \cdot m} \cdot \frac{t_{sp}}{t_{sample}} \cdot \exp{\left( \lambda \Delta t \right)},
    \label{eq:standard_pottery}
\end{equation}

\noindent where $q_{i}$ is the abundance of a chemical element in sample $i$ in ppm, $C_{i}$ is the total count of a peak produced by a delayed gamma related to an isotope of that element, $m$ is the irradiated total mass of the standard pottery, $t_i$ is the live time, $\lambda$ is the decay constant of the isotope, $\Delta t$ is the time difference between the start of the measurement of the standard pottery and the sample, $i = sample$ for one of the analyzed samples and $i = sp$ for the standard pottery.

If more than one peak was used to infer the abundance of a chemical element in a sample, a weighted average was performed. The uncertainties were obtained by error propagation of Equation \ref{eq:standard_pottery}.

Since the standard pottery was irradiated under the same flux conditions as the samples and measured with the same geometry and experimental setups, systematic uncertainties due to these conditions are eliminated.

If the isotope produced by NAA in the sample is not also present in the standard pottery, then a direct calculation was done using Equation \ref{eq:abundance}:

\begin{equation}
    m = \frac{\left( C - \Sigma_i C_i \right) M \lambda e^{- \lambda t_0}}{0.602 I \varepsilon \Phi \sigma \left(1 - e^{-\lambda t_{irr}} \right) \left(1 - e^{-\lambda \Delta t} \right) } \cdot \left( \frac{\Delta t}{\Delta t_{live}} \right)^{-1},
    \label{eq:abundance}
\end{equation}

\noindent where $t_0$ is the time between the end of the irradiation and the beginning of the measurement, $I$ is the gamma branch ratio, $\Phi$ is the flux of neutrons, $\sigma$ is the thermal neutron cross section, $M$ is the molar mass of the isotope, $t_{irr}$ is the irradiation time, $\Delta t$ is the real time of measurement and $\Delta t_{live}$ is the live time. $\left( \Delta t/\Delta t_{live} \right)^{-1} $ is the time correction due to dead time. $C$ is the total counting of the peak, $C_i$ is the expected number of events due to some contamination, if present. It was estimated by using a "pure" gamma line in the same spectra.

Equation \ref{eq:abundance} was used with the standard pottery to estimate the neutron flux. Error propagation with covariance matrix analysis was used to estimate the uncertainty of $m$.

\subsubsection{Detector Set-up}

Three main setups were used to measure the delayed gammas. One setup included an HPGe n-type detector of 2.1 kg with $85\%$ relative efficiency from ORTEC. The data were acquired with a multi-channel analyzer of 8192 channels. An anti-coincidence system was used, with 5 NaI detectors, four of 4''x4''x4'' and one of 6''x6''x6'' surrounding the samples with an HPGe p-type detector of 60\% relative efficiency  facing the sample. Each NaI detector was connected to a timing single channel analyzer (timing-SCA), which generated an output logic pulse if the input signal fell within the pulse-height window of 0 to 3 MeV. The five timing-SCAs were connected to a universal coincidence module to work as one detector and then connected to a gate and delay generator to delay the signal by less than $0.5 \mu s$. The 8192 channels MCA performed the anti-coincidence, registering only events that occurred in the HPGe detectors and NOT in the NaI detectors.

The anti-coincidence efficiency was calculated using a ${}^{60}Co$ source. The ratio of the area under the 1173 keV peak ($\mathcal{A}_p$) over the area of the Compton due to that peak ($\mathcal{A}C$) was evaluated with and without anti-coincidence, $r_I$ and $r{II}$, respectively:

\begin{equation}
\begin{split}
r_I = \frac{\mathcal{A}_p}{\mathcal{A}C} = 0.19 \\
r_{II} = \frac{\mathcal{A}_p}{\mathcal{A}_C} = 0.30
\end{split}
\end{equation}

The suppression factor, i.e., the ratio of the total counting of the spectra with and without the anti-coincidence system, $\mathcal{A}_s$, was also evaluated:

\begin{equation} \mathcal{A}_s = 4.3 \end{equation}

Lastly, the ratio of the sum-peak with and without the anti-coincidence system, $\mathcal{A}_{ps}$, was evaluated:

\begin{equation} \mathcal{A}_{ps} = 1.19 \end{equation}

A coincidence system with two HPGe detectors was set up to measure the iridium abundances by searching for delayed gammas from the decay of ${}^{192}Ir$ produced by thermal neutron absorption of ${}^{191}Ir$. 
The ${}^{192}Ir$ has four intense gamma-rays in coincidence: $E_\gamma = 296.0 , \text{keV}$ ($I_\gamma = 28.71\%$), $E_\gamma = 308.5 , \text{keV}$ ($I_\gamma = 29.70\%$), $E_\gamma = 316.5 , \text{keV}$ ($I_\gamma = 82.86\%$), and $E_\gamma = 468.1 , \text{keV}$ ($I_\gamma = 47.84\%$), so the windows of the timing-SCA were configured to operate between 289 keV and 323 keV.

The efficiency of the coincidence system was tested with a sample of $IrO_2$ of $m=1.0107 \, \text{g}$ irradiated with thermal neutrons. The 468 keV peak was used to evaluate the total efficiency. Without the coincidence system, it was $\varepsilon = 0.06870 (22)$. With it, $\varepsilon = 0.00747 (5)$, i.e., 10 times smaller.


\subsubsection{Detector Efficiency Calibration}

Calibration curves for each HPGe setup and for each sample distance were produced using ${}^{22}Na$, ${}^{57}Co$, ${}^{60}Co$, ${}^{133}Ba$, ${}^{137}Cs$, ${}^{152}Eu$, and ${}^{228}Th$ radioactive sources. These efficiency curves include intrinsic and geometric components. Each efficiency data point was calculated by $\varepsilon = C/(I \cdot A)$, where $I$ is the gamma intensity and $A$ is the activity of the source during the efficiency measurement. The activity was calculated from the nominal activity of the sample $A_0$ using the relation $A \approx A_0 \exp{\left( -\lambda t_i \right) } \delta t$, where $t_i$ is the time difference between when the nominal activity was measured and the current time, and $\Delta t$ is the measurement time. This equation is valid in the approximation $\Delta t \lambda \ll 1$.

The efficiencies were estimated by fitting the function of Equation \ref{eq:eff} to the data.

\begin{equation} \varepsilon = \frac{1}{E_\gamma} \sum\limits_{\substack{i=0 \ i \neq 4,6}}^7 p_i \ln^i \left( E_\gamma \right), \label{eq:eff} \end{equation}

\noindent where $E_\gamma$ is the energy of the gamma and $p_i$ are the free parameters. 


\section{Results}

The results for the Hawaiian and Kilimanjaro samples are presented in Table \ref{tab:Hawai_Kili}. For these samples, we performed two irradiations: one focused on measuring short half-life isotopes (Al, As, Ca, Cl, Dy, Mg, Mn, Na) and the other on long half-life isotopes, allowing us to measure 33 isotopes. ${}^{28}Al$ was the shortest half-life isotope measured ($T_{1/2} = 2.245 \, \text{min}$), demonstrating the wide application of this experimental setup. Tables \ref{tab:Ec_Ice_Ind} and \ref{tab:Et_Rwan_Ugan} present the results for the Ecuador, Iceland, India, Mt. Etna, Rwanda, and Uganda samples. The abundances ranged from $10^{-1}$ to $10^5$ ppm with samples of around 1 gram, which also demonstrates that NAA with HPGe detectors is a reliable technique.

\begin{sidewaystable}[h]
\centering
\footnotesize
\caption{Abundance of elements in volcanic rocks from Hawaii and Kilimanjaro. Values in ppm (parts per million).}
\begin{tabular}{|c|c|c|c|c|c|c|}
\hline
Element\textbackslash Sample & Kauai (ppm) & Kilauea (ppm) & Mauna Loa (ppm) & Haleakala (ppm) & Kilimanjaro 1 (ppm) & Kilimanjaro 2 (ppm) \\ \hline
mass (g) & 1.1238 & 1.0631 & 1.1408 & 1.3540 & 1.2101 & 1.2054 \\ \hline
Al & $ 7.1 (11) \cdot 10^{4}$ & $ 8.0 (7) \cdot 10^{4}$ & $ 5.69 (8) \cdot 10^{4}$ & $ 1.88 (20) \cdot 10^{5} $ & $ 9.2 (12) \cdot 10^{4} $ & $ 8.65 (11) \cdot 10^{4} $ \\ \hline
As &  --  & -- & -- & -- & $ 2.7 (5) $ & $ 2.6 (5) $ \\ \hline
Ba & $ 3.10 (6) \cdot 10^{2} $ & $ 1.09 (4) \cdot 10^{2} $ & $ 1.00 (3) \cdot 10^{2} $ & $ 4.89 (5) \cdot 10^{2} $ & $ 8.60 (10) \cdot 10^{2} $ & $ 8.77 (27) \cdot 10^{2} $ \\ \hline
Ca & $ 6.4 (15) \cdot 10^{3} $ & $ 7.5 (12) \cdot 10^{3} $ & $ 5.6 (13) \cdot 10^{3} $ & $ 4.7 (7) \cdot 10^{3} $ & $ 8 (3) \cdot 10^{2} $ & $ 1.2 (3) \cdot 10^{3} $ \\ \hline
Ce & $ 4.74 (16)\cdot 10 $ & $ 3.31 (11)\cdot 10 $ & $ 2.23 (6)\cdot 10 $ & $ 6.76 (13)\cdot 10 $ & $ 2.72 (6) \cdot 10^{2} $ & $ 3.19 (11) \cdot 10^{2} $ \\ \hline
Cl & $ 10 (6) $ & $ 2.5 (8) \cdot 10^{2} $ & --  & $ 2.0 (5) \cdot 10^{2} $ & $ 1.23 (22) \cdot 10^{3} $ & $ 1.31 (28) \cdot 10^{3} $ \\ \hline
Co & $ 6.78 (3)\cdot 10 $ & $ 4.302 (27)\cdot 10 $ & $ 5.325 (25)\cdot 10 $ & $ 5.273 (15)\cdot 10 $ & $ 3.797 (17) $ & $ 4.180 (27) $ \\ \hline
Cr & $ 1.369 (25) \cdot 10^{3} $ & $ 1.2 (3) \cdot 10 $ & $ 1.221 (23)\cdot 10^{3} $ & $ 1.222 (15)\cdot 10^{3} $ & $ 4.13 (6) \cdot 10^{-1} $ & $ 1.849 (29) \cdot 10^{-1} $ \\ \hline
Cs & $ 1.78 (7) $ & $ 8.81 (29)  \cdot 10^{-1} $ & $ 8.02 (25) \cdot 10^{-1} $ & $ 6.01 (16) \cdot 10^{-1} $ & $ 1.86 (3)\cdot 10 $ & $ 2.05 (7)\cdot 10 $ \\ \hline
Dy & $ 4.38 (26) $ & $ 6.33 (22) $ & $ 4.37 (25) $ & $ 4.47 (17) $ & $ 7.2 (3) $ & $ 8.1 (4) $ \\ \hline
Eu & $ 1.238 (5) $ & $ 1.560 (8) $ & $ 1.101 (5) $ & $ 1.676 (5) $ & $ 3.281 (9) $ & $ 4.289 (26) $ \\ \hline
Fe & $ 8.67 (3) \cdot 10^{4} $ & $ 7.70 (4) \cdot 10^{4} $ & $ 7.98 (3) \cdot 10^{4} $ & $ 8.700 (23) \cdot 10^{4} $ & $ 3.100 (10) \cdot 10^{4} $ & $ 3.384 (18) \cdot 10^{4} $ \\ \hline
Hf & $ 2.51 (6) $ & $ 3.89 (11) $ & $ 2.94 (8) $ & $ 5.14 (9) $ & $ 2.43 (5) \cdot 10 $ & $ 2.83 (11)\cdot 10 $ \\ \hline
K & $ 7.3 (15) \cdot 10^{3} $ & $ 3.3 (8) \cdot 10^{3} $ & $ 3.7 (14) \cdot 10^{3} $ & $ 2.9 (9) \cdot 10^{3} $ & $ 4.47 (6) \cdot 10^{4} $ & $ 6.1 (10) \cdot 10^{4} $ \\ \hline
La & $ 1.88 (5)\cdot 10 $ & $ 1.24 (4)\cdot 10 $ & $ 8.16 (20) $ & $ 1.90 (6)\cdot 10 $ & $ 1.82 (4) \cdot 10^{2} $ & $ 1.500 (13) \cdot 10^{2} $ \\ \hline
Lu & $ 2.35 (19) \cdot 10^{-1} $ & $ 4.0 (3)  \cdot 10^{-1} $ & $ 6.4 (6) \cdot 10^{-1} $ & $ 7.9 (8) \cdot 10^{-1} $ & $ 4.7 (24) \cdot 10^{-1} $ & $ 5.3 (5) \cdot 10^{-1} $ \\ \hline
Mg & $ 5.5 (9) \cdot 10^{3} $ & $ 3.9 (5) \cdot 10^{3} $ & $ 4.2 (8) \cdot 10^{3} $ & $ 3.9 (5) \cdot 10^{3} $ & $ 2.5 (4) \cdot 10^{3} $ & $ 3.06 (5) \cdot 10^{3} $ \\ \hline
Mn & $ 1.275 (17) \cdot 10^{3} $ & $ 1.459 (13) \cdot 10^{3} $ & $ 1.258 (17) \cdot 10^{3} $ & $ 1.430 (12) \cdot 10^{3} $ & $ 1.453 (13) \cdot 10^{3} $ & $ 1.499 (20) \cdot 10^{3} $ \\ \hline
Na & $ 1.307 (19) \cdot 10^{4} $ & $ 1.742 (17) \cdot 10^{4} $ & $ 1.353 (20) \cdot 10^{4} $ & $ 1.408 (13) \cdot 10^{4} $ & $ 6.24 (6) \cdot 10^{4} $ & $ 6.10 (9) \cdot 10^{4} $ \\ \hline
Nd & $ 1.86 (6)\cdot 10 $ & $ 1.67 (7)\cdot 10 $ & $ 1.37 (6)\cdot 10 $ & $ 2.36 (9)\cdot 10 $ & $ 1.24 (4)\cdot 10 $ & $ 3.13 (10)\cdot 10 $ \\ \hline
Rb & $ 2.37 (19)\cdot 10 $ & $ 9.6 (8) $ & $1.01 (8) \cdot 10$ & $ 1.34 (11)\cdot 10 $ & $ 1.83 (16) \cdot 10^{3} $ & $ 8.2 (7) \cdot 10 $ \\ \hline
Sb & $ 8.33 (16) \cdot 10^{-2} $ & $ 4.79 (6)  \cdot 10^{-1} $ & $ 2.074 (24)  \cdot 10^{-1} $ & $ 2.301 (24) \cdot 10^{-1} $ & $ 5.58 (4) \cdot 10^{-1} $ & $ 5.90 (14) \cdot 10^{-1} $ \\ \hline
Sc & $ 2.590 (23)\cdot 10 $ & $ 3.16 (4)\cdot 10 $ & $ 2.982 (27)\cdot 10 $ & $ 2.189 (13)\cdot 10 $ & $ 1.228 (9) $ & $ 1.385 (16) $ \\ \hline
Sm & $ 7.0 (13) $ & $ 7.8 (7) $ & $ 5.0 (8) $ & $ 9.2 (14) $ & $ 1.48 (17)\cdot 10 $ & $ 1.57 (16)\cdot 10 $ \\ \hline
Sr & $ 4.5 (4) \cdot 10^{2} $ & $ 3.7 (4) \cdot 10^{2} $ & $ 2.98 (26) \cdot 10^{2} $ & $ 4.80 (28) \cdot 10^{2} $ & $ 4.9 (3) \cdot 10^{2} $ & $ 2.20 (15) \cdot 10^{2} $ \\ \hline
Ta & $ 1.581 (9) $ & $ 8.60 (7)  \cdot 10^{-1} $ & $ 4.58 (4)  \cdot 10^{-1} $ & $ 2.111 (8) $ & $ 1.743 (7)\cdot 10 $ & $ 2.185 (22)\cdot 10 $ \\ \hline
Tb & $ 6.24 (9)  \cdot 10^{-1} $ & $ 9.07 (15)  \cdot 10^{-1} $ & $ 7.00 (9) \cdot 10^{-1} $ & $ 7.77 (9) \cdot 10^{-1} $ & $ 4.65 (5) \cdot 10^{-1} $ & $ 2.59 (4) \cdot 10^{-1} $ \\ \hline
Th & $ 2.300 (18) $ & $ 1.030 (14) $ & $ 6.81 (7)  \cdot 10^{-1} $ & $ 3.590 (18) $ & $ 3.725 (20)\cdot 10 $ & $ 4.13 (6)\cdot 10 $ \\ \hline
Ti & $1.25 (7) \cdot 10^{4}$ & $1.61 (8)  \cdot 10^{4}$ & $1.01 (4)  \cdot 10^{4}$ & $1.28 (7)  \cdot 10^{4}$ & $ 5.2 (6)\cdot 10^3 $ & $ 4.6 (3)\cdot 10^3 $ \\ \hline
U & $ 3.2 (10)  \cdot 10^{-1} $ & $ 4.8 (21)  \cdot 10^{-1} $ & -- & $ 7.5 (13) \cdot 10^{-1} $ & $ 8.7 (5) $ & $ 3.89 (20) $ \\ \hline
Yb & $ 1.27 (9) $ & $ 2.07 (17) $ & $ 1.67 (14) $ & $ 1.67 (6) $ & $ 3.98 (12) $ & $ 3.3 (3) $ \\ \hline
Zn & $ 2.37 (8) \cdot 10^{2} $ & $ 3.13 (22) \cdot 10^{2} $ & $ 1.96 (7) \cdot 10^{2} $ & $ 1.63 (4) \cdot 10^{2} $ & $ 2.16 (6) \cdot 10^{2} $ & $ 2.25 (10) \cdot 10^{2} $ \\ \hline
Zr & $ 1.89 (18) \cdot 10^{2} $ & $ 1.7 (4) \cdot 10^{2} $ & $1.27 (28) \cdot 10^{2}$ & $ 2.25 (25) \cdot 10^{2} $ & $ 1.33 (4) \cdot 10^{3} $ & $ 7.87 (20) \cdot 10^{2} $ \\ \hline
\end{tabular}
\label{tab:Hawai_Kili}
\end{sidewaystable}

\begin{sidewaystable}[h]
\centering
\footnotesize
\caption{Abundance of elements determined in volcanic rocks in samples from Ecuador, Iceland, and India. Values in ppm (parts per million).}
\begin{tabular}{|c|c|c|c|c|c|c|}
\hline
Element\textbackslash Sample & Ecuador 1 (ppm) & Ecuador 2 (ppm) & Iceland 1 (ppm) & Iceland 2 (ppm) & India 2 (ppm) & India Basalt (ppm)\\ \hline
mass (g) & $ 0.6827$ & $ 2.1431$ & $  1.2369$ & $ 2.1238$ & $ 2.53$ & $  2.1053$ \\ \hline
Ba & $ 1.05(4) \cdot 10^{3}$ & $ 2.96(18) \cdot 10^{2}$ & $ 1.50(13) \cdot 10^{2}$ & $ 1.4(3) \cdot 10^{2}$ & $ 7.4(4) \cdot 10^{2}$ & $ 7.3(11) \cdot 10$ \\ \hline
Ce & $ 2.57(9) \cdot 10$ & $ 1.42(6) \cdot 10$ & $ 4.62(16) \cdot 10$ & $ 4.05(14) \cdot 10$ & $ 3.31(12) \cdot 10$ & $ 3.13(11) \cdot 10$ \\ \hline
Co & $ 1.503(13) \cdot 10$ & $ 2.148(19) \cdot 10$ & $ 3.391(27) \cdot 10$ & $ 5.00(4) \cdot 10$ & $ 3.112(26) \cdot 10$ & $ 4.89(4) \cdot 10$ \\ \hline
Cr & $ 1.041(26) \cdot 10^{2}$ & $ 6.49(17) \cdot 10$ & $ 4.98(12) \cdot 10^{2}$ & $ 3.27(8) \cdot 10^{2}$ & $ 3.43(8) \cdot 10^{2}$ & $ 7.44(19) \cdot 10$ \\ \hline
Cs & $ 7.0(3) $ & $ 6.9(4) $ & $ 1.36(12) $ & $ 9.0(8) \cdot 10^{-1}$ & $ 8.7(4) $ & $ 3.40(17) $ \\ \hline
Eu & $ 7.37(8) \cdot 10^{-1}$ & $ 8.12(10) \cdot 10^{-1}$ & $ 1.925(19) $ & $ 1.876(20) $ & $ 1.536(15) $ & $ 1.930(20) $ \\ \hline
Fe & $ 6.25(4) \cdot 10^{4}$ & $ 3.467(25) \cdot 10^{4}$ & $ 1.126(8) \cdot 10^{5}$ & $ 7.52(5) \cdot 10^{4}$ & $ 9.02(6) \cdot 10^{4}$ & $ 1.490(10) \cdot 10^{5}$ \\ \hline
Hf & $ 4.36(19) $ & $ 2.26(20) $ & $ 6.7(3) $ & $ 3.67(17) $ & $ 3.00(16) $ & $ 5.96(27) $ \\ \hline
K & $ 1.7(5) \cdot 10^{4}$ & $ 6.0(4) \cdot 10^{3}$ & $ 1.1(4) \cdot 10^{4}$ & $ 4.40(25) \cdot 10^{3}$ & $ 1.5(28) \cdot 10^{3}$ & $ 5(3) \cdot 10^{3}$ \\ \hline
La & $ 3.15(27) \cdot 10^{2}$ & $ 7.8(13) \cdot 10$ & $ 7.9(6) \cdot 10$ & $ 1.1(3) \cdot 10$ & $ 6.5(3) $ & $ 1.93(9) $ \\ \hline
Lu & $ 3.8(4) \cdot 10^{-1}$ & $ 4.8(5) \cdot 10^{-1}$ & $ 4.3(4) \cdot 10^{-1}$ & $ 1.35(11) \cdot 10^{-1}$ & $ 1.95(20) \cdot 10^{-1}$ & $ 2.36(23) \cdot 10^{-1}$ \\ \hline
Nd & $ 1.2(4) \cdot 10$ & $ 5.6(10) \cdot 10$ & $ 3.1(5) \cdot 10$ & $ 2.8(4) \cdot 10$ & $ 1.7(5) $ & $ 2.0(3) \cdot 10$ \\ \hline
Rb & $ 4.0(3) \cdot 10$ & $ 2.6(3) \cdot 10$ & $ 2.00(20) \cdot 10$ & $ 2.4(7) \cdot 10$ & $ 2.64(29) \cdot 10$ & $ 9.2(13) $ \\ \hline
Sb & $ 4.67(20) \cdot 10^{-1}$ & $ 2.71(28) \cdot 10^{-1}$ & $ 1.88(12) \cdot 10^{-1}$ & -- & $ 9.28(19) $ & -- \\ \hline
Sc & $ 1.788(29) \cdot 10$ & $ 1.43(23) \cdot 10$ & $ 2.82(4) \cdot 10$ & $ 2.245(26) \cdot 10$ & $ 2.59(4) \cdot 10$ & $ 4.59(7) \cdot 10$ \\ \hline
Sr & $ 5.5(7) \cdot 10^{2}$ & $ 3.9(5) \cdot 10^{2}$ & $ 4.3(5) \cdot 10^{2}$ & $ 3.7(4) \cdot 10^{2}$ & $ 7.5(9) \cdot 10^{2}$ & $ 2.9(4) \cdot 10^{2}$ \\ \hline
Ta & $ 1.75(8) \cdot 10^{-1}$ & $ 1.59(12) \cdot 10^{-1}$ & $ 1.652(26) $ & $ 1.075(16) $ & $ 2.14(7) \cdot 10^{-1}$ & $ 5.43(13) \cdot 10^{-1}$ \\ \hline
Tb & $ 4.12(24) \cdot 10^{-1}$ & $ 3.52(26) \cdot 10^{-1}$ & $ 9.1(5) \cdot 10^{-1}$ & $ 7.6(5) \cdot 10^{-1}$ & $ 5.2(4) \cdot 10^{-1}$ & $ 1.23(9) $ \\ \hline
Th & $ 2.73(6) $ & $ 8.1(3) \cdot 10^{-1}$ & $ 3.02(7) $ & $ 1.49(4) $ & $ 5.50(12) $ & $ 3.65(8) $ \\ \hline
U & -- & -- & -- & -- & -- & -- \\ \hline
Yb & $ 1.89(18) $ & $ 6.6(8) \cdot 10^{-1}$ & $ 2.55(24) $ & $ 1.86(17) $ & $ 2.12(20) $ & $ 4.8(4) $ \\ \hline
Zn & $ 1.47(7) \cdot 10^{2}$ & $ 1.87(9) \cdot 10^{2}$ & $ 1.37(6) \cdot 10^{2}$ & $ 1.18(6) \cdot 10^{2}$ & $ 8.6(4) \cdot 10$ & $ 2.53(16) \cdot 10^{2}$ \\ \hline
Zr & $ 5.19(25) \cdot 10$ & $ 3.7(7) \cdot 10$ & $ 1.10(4) \cdot 10^{2}$ & $ 8.8(3) \cdot 10$ & $ 7.1(4) \cdot 10$ & $ 6.9(3) \cdot 10$ \\ \hline
\end{tabular}
\label{tab:Ec_Ice_Ind}
\end{sidewaystable}

\begin{sidewaystable}[h]
\centering
\footnotesize
\caption{Abundance of elements determined in volcanic rocks in samples from Mt. Etna, Rwanda, and Uganda. Values in ppm (parts per million).}
\begin{tabular}{|c|c|c|c|c|c|}
\hline
Element\textbackslash Sample & Mt. Etna 1 (ppm) & Mt. Etna 2 (ppm) & Rwanda (ppm) & Uganda 1 (ppm) & Uganda 2 (ppm)\\ \hline
mass (g) & $ 2.011$ & $  1.1740$ & $ 2.0186$ & $ 1.7729$ & $ 1.2928$ \\ \hline
Ba & $ 3.14(22) \cdot 10^{2}$ & $ 1.23(6) \cdot 10^{3}$ & $ 2.76(12) \cdot 10^{3}$ & $ 7.7(3) \cdot 10^{2}$ & $ 1.20(6) \cdot 10^{3}$ \\ \hline
Ce & $ 9.8(3) \cdot 10$ & $ 1.15(4) \cdot 10^{2}$ & $ 2.68(9) \cdot 10^{2}$ & $ 1.48(5) \cdot 10^{2}$ & $ 1.86(6) \cdot 10^{2}$ \\ \hline
Co & $ 3.576(29) \cdot 10$ & $ 3.404(28) \cdot 10$ & $ 1.250(10) \cdot 10$ & $ 6.64(5) \cdot 10$ & $ 3.88(3) \cdot 10$ \\ \hline
Cr & $ 4.45(11) \cdot 10$ & $ 1.94(5) \cdot 10$ & $ 6.01(19) $ & $ 1.52(4) \cdot 10^{3}$ & $ 3.02(7) \cdot 10^{2}$ \\ \hline
Cs & $ 5.21(26) $ & $ 7.5(3) $ & $ 8.6(4) $ & $ 4.96(25) $ & $ 6.1(3) $ \\ \hline
Eu & $ 2.622(25) $ & $ 2.680(24) $ & $ 3.54(3) $ & $ 2.340(23) $ & $ 2.857(24) $ \\ \hline
Fe & $ 5.81(4) \cdot 10^{4}$ & $ 8.13(6) \cdot 10^{4}$ & $ 6.26(4) \cdot 10^{4}$ & $ 1.056(8) \cdot 10^{5}$ & $ 8.56(6) \cdot 10^{4}$ \\ \hline
Hf & $ 3.81(18) $ & $ 4.82(21) $ & $ 9.1(4) $ & $ 6.55(30) $ & $ 6.32(28) $ \\ \hline
K & $ 4(4) \cdot 10^{3}$ & $ 1.2(3) \cdot 10^{4}$ & $ 1.04(26) \cdot 10^{4}$ & $ 1.1(5) \cdot 10^{4}$ & $ 1.3(5) \cdot 10^{4}$ \\ \hline
La & $ 5.7(4) \cdot 10^{2}$ & $ 2.44(19) \cdot 10$ & $ 1.87(12) \cdot 10^{2}$ & $ 1.90(18) \cdot 10$ & $ 1.47(10) \cdot 10^{3}$ \\ \hline
Lu & $ 3.49(29) \cdot 10^{-1}$ & $ 2.99(20) \cdot 10^{-1}$ & $ 4.7(4) \cdot 10^{-1}$ & $ 2.00(18) \cdot 10^{-1}$ & $ 8.1(5) \cdot 10^{-1}$ \\ \hline
Nd & $ 7.7(10) \cdot 10$ & $ 5.1(6) \cdot 10$ & $ 1.11(14) \cdot 10^{2}$ & $ 5.1(8) \cdot 10$ & $ 1.28(16) \cdot 10^{2}$ \\ \hline
Rb & $ 2.75(21) \cdot 10$ & $ 6.6(4) \cdot 10$ & $ 2.29(15) \cdot 10^{2}$ & $ 7.9(6) \cdot 10$ & $ 1.14(8) \cdot 10^{2}$ \\ \hline
Sb & $ 1.7(4) \cdot 10^{-1}$ & $ 6.7(14) \cdot 10^{-2}$ & $ 1.16(14) \cdot 10^{-1}$ & -- & -- \\ \hline
Sc & $ 2.65(4) \cdot 10$ & $ 2.560(24) \cdot 10$ & $ 7.64(7) $ & $ 3.41(5) \cdot 10$ & $ 2.962(28) \cdot 10$ \\ \hline
Sr & $ 1.02(12) \cdot 10^{3}$ & $ 1.18(14) \cdot 10^{3}$ & $ 1.16(14) \cdot 10^{3}$ & $ 8.3(10) \cdot 10^{2}$ & $ 9.2(11) \cdot 10^{2}$ \\ \hline
Ta & $ 1.884(29) $ & $ 1.852(27) $ & $ 7.07(11) $ & $ 4.76(8) $ & $ 5.60(8) $ \\ \hline
Tb & $ 6.0(4) \cdot 10^{-1}$ & $ 1.04(6) $ & $ 1.36(8) $ & $ 4.8(4) \cdot 10^{-1}$ & $ 1.03(6) $ \\ \hline
Th & $ 5.60(10) $ & $ 8.71(13) $ & $ 3.32(4) \cdot 10$ & $ 1.387(24) \cdot 10$ & $ 1.375(21) \cdot 10$ \\ \hline
U & $ 2.0(5) $ & -- & $ 3.6(7) $ & -- & -- \\ \hline
Yb & $ 1.40(13) $ & $ 2.47(23) $ & $ 2.75(25) $ & $ 2.02(20) $ & $ 2.28(21) $ \\ \hline
Zn & $ 1.27(6) \cdot 10^{2}$ & $ 1.79(11) \cdot 10^{2}$ & $ 2.19(8) \cdot 10^{2}$ & $ 1.77(8) \cdot 10^{2}$ & $ 1.81(8) \cdot 10^{2}$ \\ \hline
Zr & $ 1.10(5) \cdot 10^{2}$ & $ 1.01(4) \cdot 10^{2}$ & $ 1.97(6) \cdot 10^{2}$ & $ 1.14(5) \cdot 10^{2}$ & $ 1.49(6) \cdot 10^{2}$ \\ \hline
\end{tabular}
\label{tab:Et_Rwan_Ugan}
\end{sidewaystable}

Several hundred peaks were observed for each of these measurements, with the majority  of them identified. A careful analysis was conducted to identify possible contamination due to pile-ups, sum, and pair production. Only pure peaks or peaks with minimal contamination were used whenever possible. When impurities when observed, they were subtracted using a pure peak of the contaminant as reference. All the peaks were manually fitted using IDEFIX software, which showed the best discrimination power in cases of multiplets \cite{Idefix}. The fits were done with a Gaussian distribution for each peak, plus a second-degree polynomial function for the background and an exponential left tail to account for HPGe charge collection effects. 

In Appendix~\ref{app1}, an example of the full spectra obtained for Kilimanjaro is presented. 


\subsection{Uranium fission and Lanthanum Estimation}

Thermal neutrons can induce fission in heavy isotopes, like ${}^{235}U$ ($\sigma = 585 \, b$), which was actually noticed by the univocal identification of ${}^{140}Ba$ peaks ($T_{1/2} = 12.8 \, d$), which can only be produced by this fission. Since ${}^{140}Ba$ decays to ${}^{140}La$, which was used to determine the abundance of lanthanum, a correction was made to account for this contribution.

\subsection{Cerium}

Another special case is the determination of the abundance of cerium, which was possible only from ${}^{141}Ce$ ($T_{1/2} = 32.5 , d$). The main difficulty for this measurement is due to the presence of only one high-intensity gamma, $E_\gamma = 145.4 , \text{keV}$, $I_\gamma = 48.3 \%$. As seen in Figure \ref{espectro Kili1_2 0}, there are hundreds of peaks nearby in that region. A careful fit with IDEFIX was done to ensure convergence. Following the peak-decay allowed us to confirm that the 145 keV peak comes from this decay.

\subsection{Iridium}

Iridium can be primarily measured by neutron activation of \({}^{191}\text{Ir}\), which has a thermal neutron cross-section of 954 b. The isotope \({}^{192}\text{Ir}\) (\(T_{1/2} = 73.8\) days) can be observed through the following gamma emissions: 612.5 keV (5.3\%), 588.6 keV (4.5\%), 468.1 keV (47.8\%), 316.5 keV (82.9\%), 308.5 keV (29.7\%), and 296.0 keV (28.7\%). Due to the presence of several highly intense peaks in this region, along with Compton scattering, no identifiable peaks were observed.

\subsection{Lutetium}

The lutetium abundance was calculated by measuring the delayed gamma from ${}^{177}Lu$ ($T_{1/2} = 160.4 \, d$). However, this isotope is also produced by the decay of ${}^{177}Yb$ ($T_{1/2} = 1.9 \, h$), which is produced by thermal capture on ${}^{176}Yb$. Since the measurements were performed one week after the irradiation, it is not possible to measure directly the decay of ${}^{176}Yb$. The peaks from ${}^{169}Yb$ ($T_{1/2} = 32.0 \, d$) were used in order to calculate the production of ${}^{177}Yb$, and we quantified a contribution of less than $< 1/1000$ to ${}^{177}Lu$.

\subsection{Thorium}
Thorium was determined from the delayed gammas of ${}^{233}Pa$ ($T_{1/2} = 27.0 \, d$), which is produced by the decay of ${}^{233}Th$ ($T_{1/2} = 21.8 \, \text{min}$). This isotope does not emit intense gammas, except for the one at 7.2 keV which was not possible to identify in our setup. 

\subsection{Zinc}
Zinc has only one intense gamma at 1115.5 keV. The main difficulty is due to the presence of high-intensity peaks close to it, like the 1112 and 1115 keV peaks emitted by ${}^{152}Eu$ and 1121 keV peak emitted by the decay of ${}^{182}Ta$ and ${}^{46}Sc$. Careful fit constraints and half-life measurements allowed us to separate this peak from the others.

\subsection{Strontium}
Strontium was identified by the gamma of ${}^{85}Sr$ ($E_\gamma = 514 \, \text{keV}$, $T_{1/2} = 64.8 \, \text{d}$\, $I_\gamma = 96 \%$). Due to the proximity of the 511 keV pair production peak, this analysis was affected. Our experimental setup was able to separate peaks of less than 2 keV, which was important in this case. 

\subsection{Chromium}
The abundance of chromium was calculated by ${}^{51}Cr$ ($T_{1/2} = 27.7 , \text{d}$), which has only one intense peak at 320.1 keV ($I_\gamma = 9.9 \%$). This peak has contamination from the 319.4 keV gamma from ${}^{147}Nd$ ($T_{1/2} = 11.0 \, \text{d}$, $I_\gamma = 2.13 \%$), which accounts for $<5 \%$ of the peak. This interference was subtracted. Additionally, a 20 times more intense 312 keV peak from ${}^{233}Pa$ was noticed but was successfully separated by the fit performed.

\subsection{Zirconium}
The most difficult case is the one of zirconium, due to the presence of several contamination. The abundance was determined with the 724.2 keV ($I_\gamma = 44.23\%$) and 756.7 keV ($I_\gamma = 54.38 \%$) gammas from ${}^{95}Zr$ ($T_{1/2} = 64.0 , \text{d}$). ${}^{154}Eu$ ($T_{1/2} = 8.6 , \text{y}$) has two intense peaks with almost the same energies as those from ${}^{95}Zr$, 723.3 keV ($I_\gamma = 20.06 \%$) and 756.8 keV ($I_\gamma = 4.52 \%$). Since ${}^{154}Eu$ has several other pure peaks, they were used to subtract this contamination from the zirconium peaks. ${}^{95}Zr$ decays to ${}^{95}Nb$ ($T_{1/2} = 3.6 , \text{d}$), which has a beta-delayed gamma at 765.8 keV ($I_\gamma = 99.8 \%$). This peak has interference from the 765.3 keV gamma from ${}^{160}Tb$ ($T_{1/2} = 72.3 , \text{d}$), whose contribution was subtracted. Since the thermal neutron fission of ${}^{235}U$ produces isotopes distributed around isotope masses of A=90 and A=140, this contribution must be accounted for. We used the thermal neutron fission distribution to account for this effect \citep{denisov2022fission}, which corresponds to an interference of $\sim 5\%$.

\subsection{Anti-coincidence Measurements}
\begin{sidewaysfigure} [h]
    \centering
    \includegraphics[width=21cm]{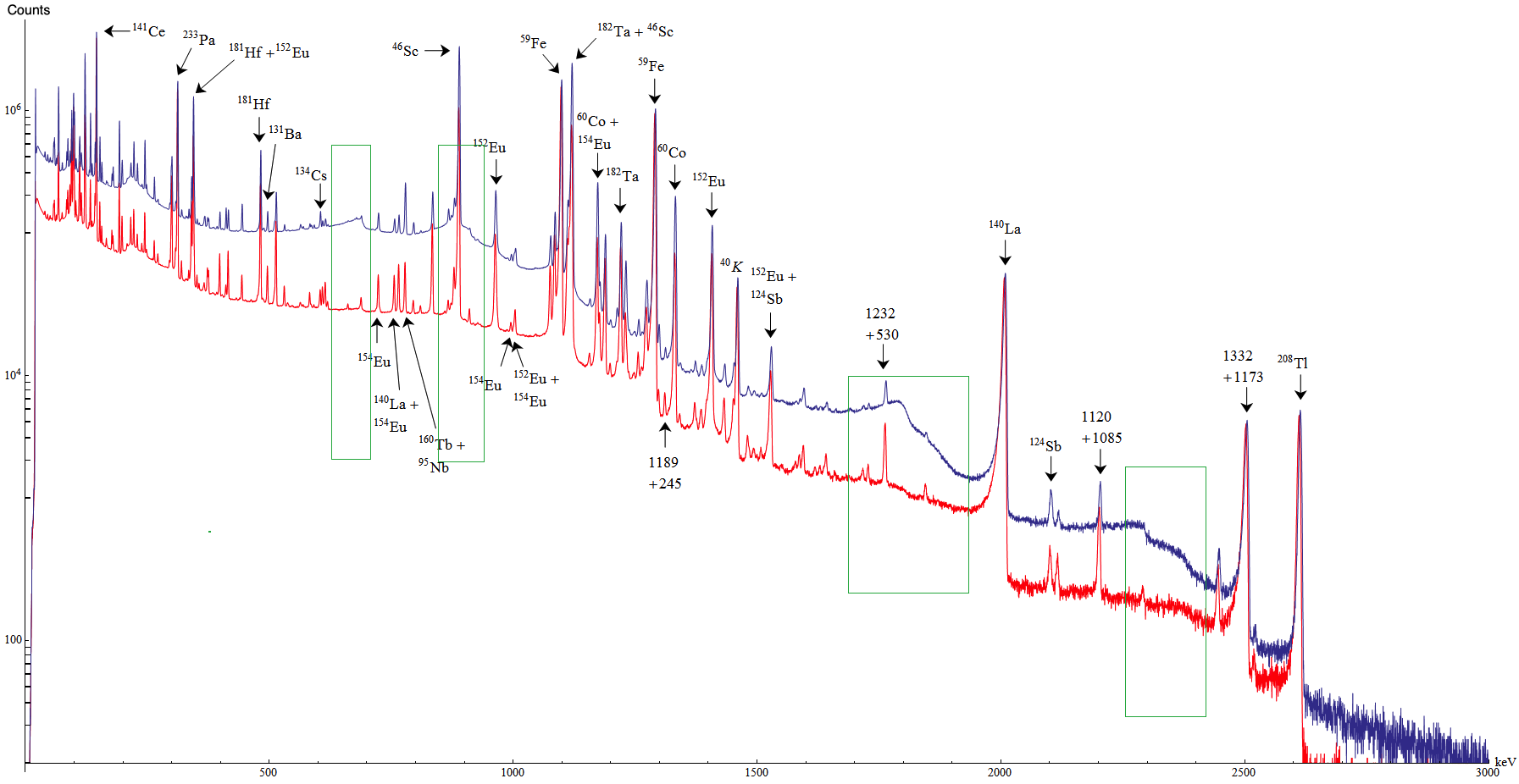}
    \caption[]{Simple and anti-coincidence spectra overlaid for Rwanda 51 days after irradiation. Live time of $72,7 \; h$ for both with the sample facing the HPGe detector. The anti-coincidence set-up reduces significantly the background, increasing the precision of the peaks. However, no peaks corresponding to new chemical elements were reported. Green box rectangles correspond to Compton regions whose backgrounds were more significantly reduced.}
    \label{fig:RwandaAnti2}
\end{sidewaysfigure}

The anti-coincidence spectra overlaid with the single spectra can be seen in Figure \ref{fig:RwandaAnti2}. Although the background, mainly due to Compton scattering, was drastically reduced, no new isotopes were identified. We specifically searched for gold from the beta-delayed gammas of ${}^{198}Au$ ($T_{1/2} = 2.7 , \text{d}$), which has one intense gamma at 411 keV, $I_\gamma = 2.2 \%$. ${}^{152}Eu$ ($T_{1/2} = 13.5 , \text{y}$) also has a gamma at 411 keV ($I_\gamma = 2.2 \%$). Even with a half-life of several years, the high abundance of ${}^{152}Eu$ prevented us from having the sensitivity to identify gold, even with the anti-coincidence setup. Comparisons among the different gammas of ${}^{152}Eu$ were done to find possible contamination on the 411 keV peak that could indicate the presence of gold, but nothing was statistically significant was observed.

\subsection{Coincidence Measurements}

\begin{figure}[t]
  \centering
  \includegraphics[scale=0.4]{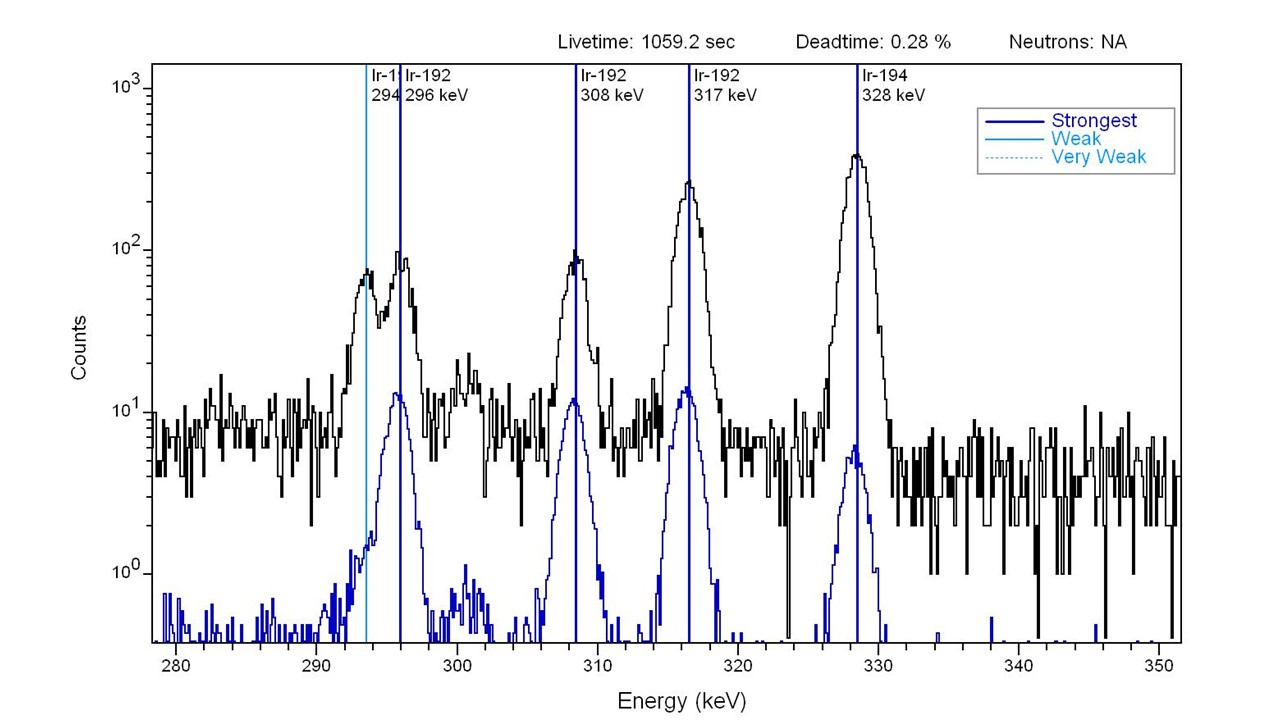}
  \caption{Spectra of a $^{192}Ir$ calibration with (blue line) and without (black line) the coincidence acquisition system. A
gate circuit was used to gate on the cluster of the $^{192}Ir$ peak from 289 – 323 keV to minimize the continuum in that
region and enhance the sensitivity for the 468 keV line. A
calibration sample of $^{192}Ir$ produced by irradiating $IrO_2$ with
neutrons was used to determine the coincidence counting
efficiency.}
   \label{fig:IrPlot}
\end{figure}

\begin{figure}[t]
  \centering
  \includegraphics[scale=0.50]{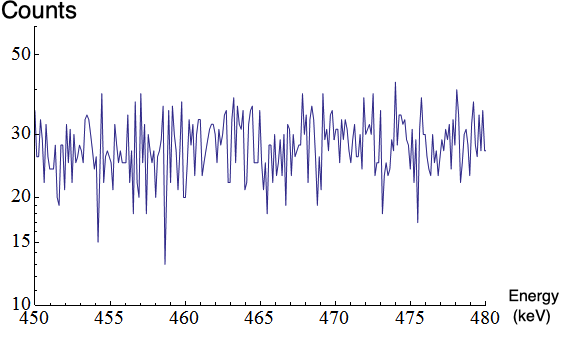}
  \caption{468 keV region from India sample 1 in coincidence with the single channel analyzer gate set from 289 – 323 keV as described in the text. }
   \label{fig:IrEcuador1}
\end{figure}

The measurements in coincidence did not allow us to identify the presence of iridium in the samples. In Figure~\ref{fig:IrEcuador1}, the spectra of India 1 is zoomed in the region of 468 keV. A peak in that region would indicate the presence of iridium in the samples. Assuming a linear background, a Bayesian fit was performed to determine the upper limit of the peak area \citep{Helene1983,Helene1991}. In Table~\ref{tab:Ir}, the upper limits of the abundance of iridium in each sample are presented. We report $<0.69 \, \text{ppb}$ in the samples. In Figures~\ref{fig:IrPlot} there is a spectra of a IrO$_2$ sample irradiated with neutrons to evaluate the efficiency of the coincidence system. The coincidence set-up successfully suppress background, while keeping the coincidence peaks. 


\begin{table}[t]
\centering
\caption{Upper limits established for Ir in our samples.}
\begin{tabular}[c]{|c|c|}
\hline
Origin & MLD (ppb)\\ \hline
Ecuador 1 & $ <0.59 $\\ \hline
Ecuador 2 & $ <0.40 $\\ \hline
Iceland 1 & $ <0.69  $\\ \hline
Iceland 2 & $ <0.29 $\\ \hline
India 1 & $ <0.62 $\\ \hline
India 2 & $ <0.38 $\\ \hline
Mt. Etna 1 & $ <0.43 $\\ \hline
Mt. Etna 2 & $ <0.52 $\\ \hline
Rwanda & $ <0.38 $\\ \hline
Uganda 1 & $ <0.51 $\\ \hline
Uganda 2 & $ <0.64 $\\ \hline
\end{tabular} 
\label{tab:Ir}
\end{table}

\subsection{Hawaii's Archipelago Analysis}

\begin{figure}[h] \centering \includegraphics[scale=0.30]{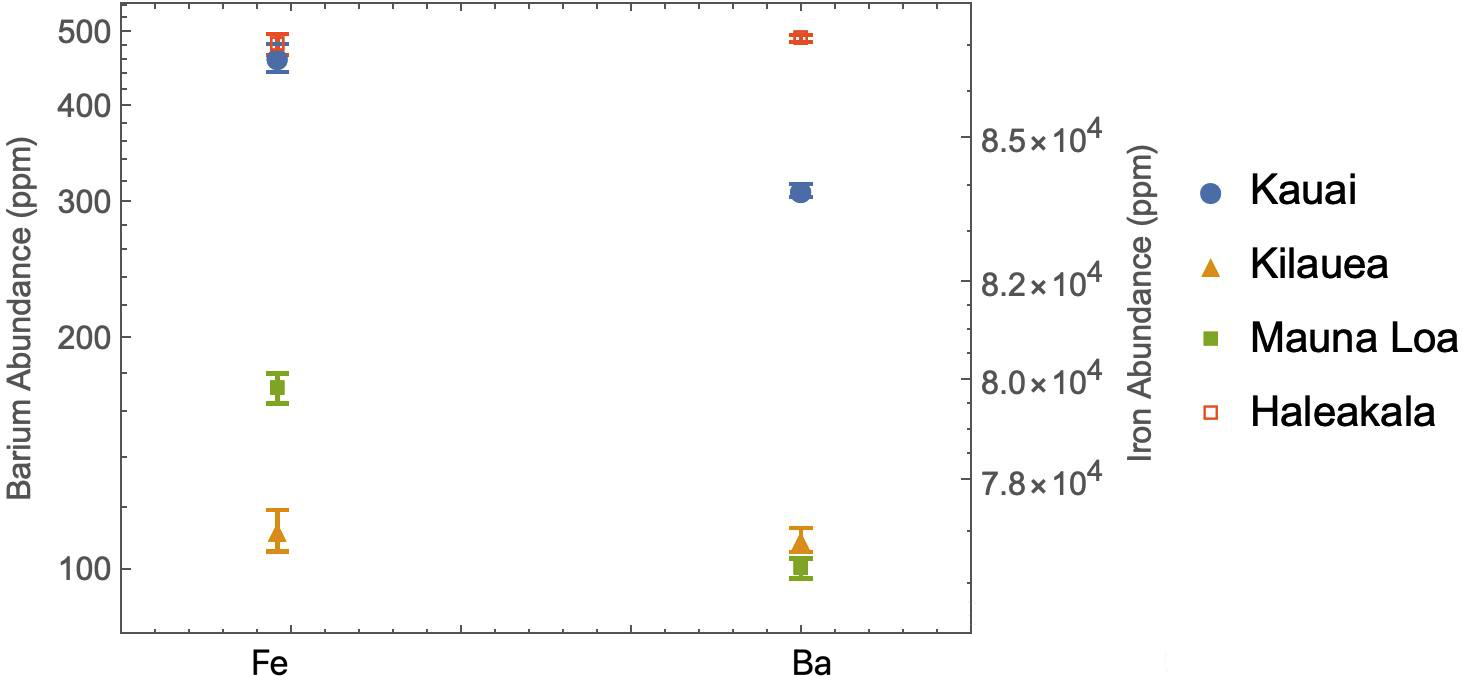} \caption[Comparison of Hawaiian samples.] {Comparison of Hawaiian samples.} \label{fig:hawaii_comp} \end{figure}

Lastly, we analyzed all the abundances of the Hawaiian samples. Only barium and iron abundances have uncertainties small enough among the samples, as presented in Figure~\ref{fig:hawaii_comp}, which allows to notice a possible correlated pattern. From this, we obtained a positive correlation between the Kilauea and Mauna Loa samples, which is expected since they both come from the same island, and between the Kauai and Haleakala samples.

\section{Conclusion}

Neutron activation analysis with gamma spectroscopy done by HPGe detectors in different configurations has been proven to be a simple but powerful technique to measure the abundances of elements in volcanic samples, with sensitivity ranging from $10^5$ ppm to ppb. With a systematic study of all gamma lines identified in the spectra, contamination subtraction by covariance analysis, measurement of the half-life from each peak, and powerful fit techniques, even the most difficult isotopes, such as zirconium, uranium, and lanthanum, could be identified and quantified.

Using this hardware-software setup, it was possible to identify 33 chemical elements, including 21 trace ones, 20 heavy metals, and 9 rare earth elements: Al, As, Ba, Ca, Ce, Cl, Co, Cr, Cs, Dy, Eu, Fe, Hf, K, La, Lu, Mg, Mn, Na, Nd, Rb, Sb, Sc, Sm, Sr, Ta, Tb, Th, Ti, U, Yb, Zn, and Zr.

The results obtained allowed us to set stringent limits on the abundance of iridium in those samples (<0.69 ppb), excluding volcanic activity as the main cause of the Cretaceous-Tertiary Mass Extinction. Additionally, we were able to measure inhomogeneities in the abundance of all samples. Specifically, uranium and thorium were measured in the range of $10$ to $10^{-1}$ ppm, which can be used to update geoneutrino emission simulations, reducing uncertainties in their abundance in the inner core. The abundance of neodymium measured in Rwanda [$111(14) \; \text{ppm}$] and Uganda 1 [$128 (16) \; \text{ppm}$] samples, and of lanthanum [$1470 (10) \; \text{ppm}$] in the Uganda 2 sample, shows a potential interest in exploring volcanic sites as a source of strategic rare earth elements.

Although not completely proven, the measurement of barium and iron samples from the Hawaiian Archipelago is compatible with the hypothesis of two lava sources originating it. We indicate that a simple high-precision NAA with a correlation analysis of the abundance of the chemical elements could provide substantial information to address this puzzle, without the need for isotopic measurements.

Despite the challenges in separating gamma lines from high abundance to trace ones, this work demonstrates that it is feasible to build a flexible setup to obtain a comprehensive picture of the abundances of elements in a sample, including trace ones, with minimal hardware modifications. Using this approach, we systematically studied 17 volcanic samples. These measurements can be used to understand magma dynamics and mantle inhomogeneity, together with other measurements and simulations.

\section{Acknowledgements}
This work was partially supported by the U. S. Department of Energy National Nuclear Security Administration under Award Number DE- NA0000979.

\section{Author contributions}
P.V.G., E.B.N., K.J.T., and A.R.S. performed the sample preparations, neutron activations, detector setups, and gamma-ray counting.  All of the authors participated in the data analysis.  The paper was written by P.V.G., E.B.N. and K.J.T.

\clearpage
\appendix
\section{Spectra for Kilimanjaro Samples}
\label{app1}

In this Appendix it is presented a full spectra of the Kilimanjaro sample, with the majority of the peaks labeled.

\begin{sidewaysfigure}[h]
    \centering
    \includegraphics[width=21cm]{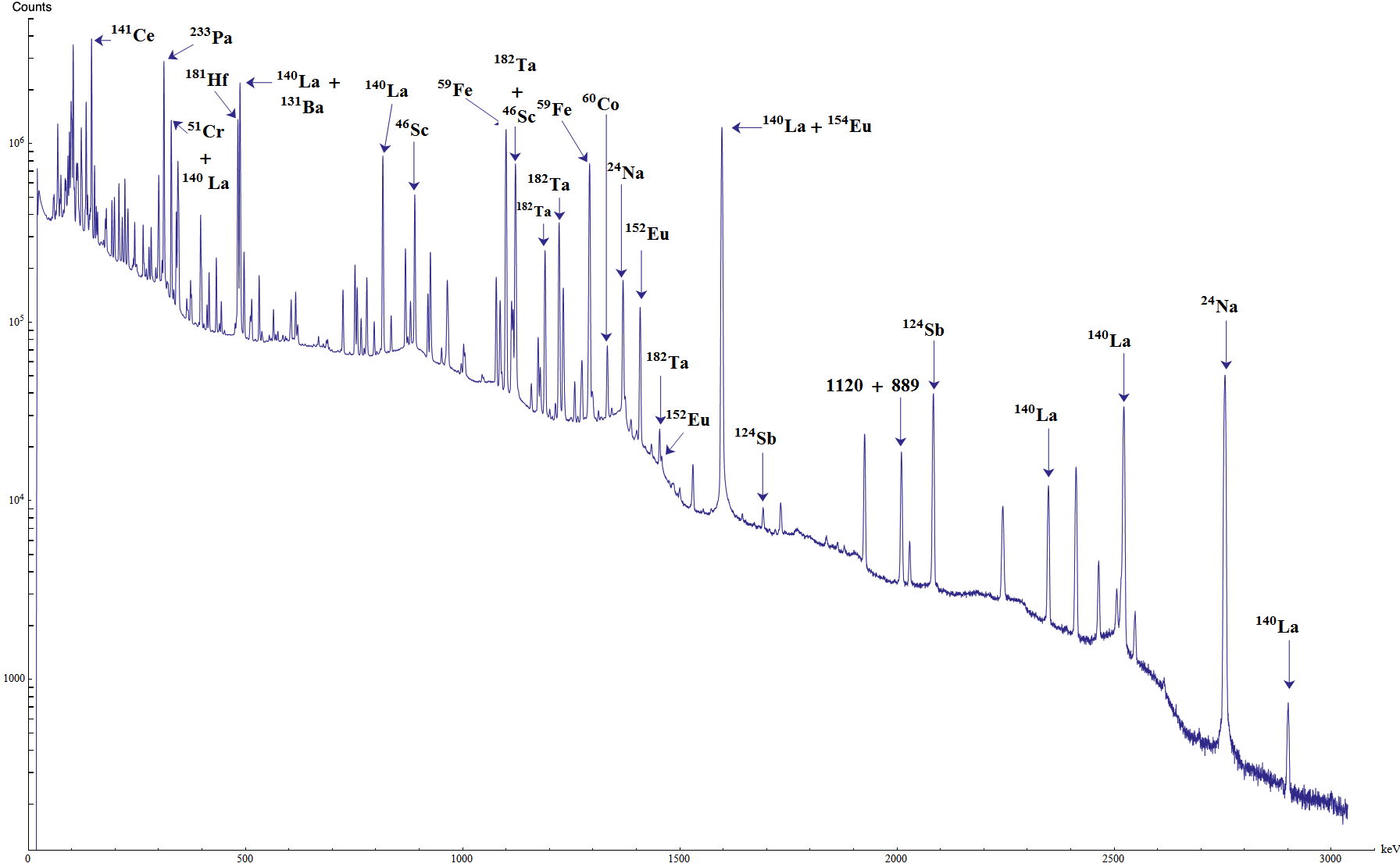}
    \caption[]{Full spectrum of one of the Kilimanjaro samples with most peaks identified.
    }
    \label{espectro inteiro Kili1_2}
\end{sidewaysfigure}

\begin{sidewaysfigure}[h] 
    \centering
    \includegraphics[width=21cm]{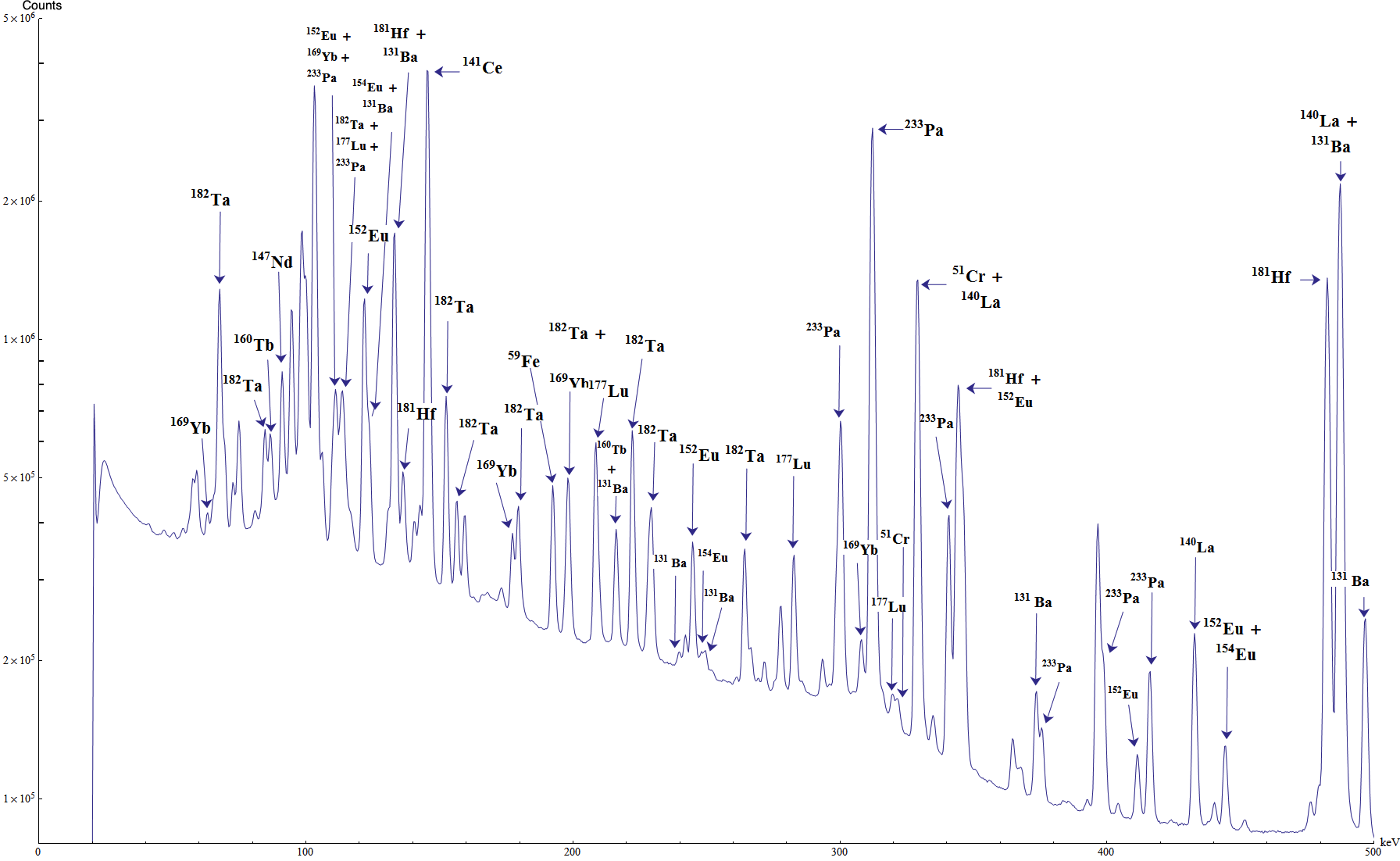}
    \caption[]{Full spectrum of one of the Kilimanjaro samples with most peaks identified.
    }
    \label{espectro Kili1_2 0}
\end{sidewaysfigure}

\begin{sidewaysfigure} [h]
    \centering
    \includegraphics[width=21cm]{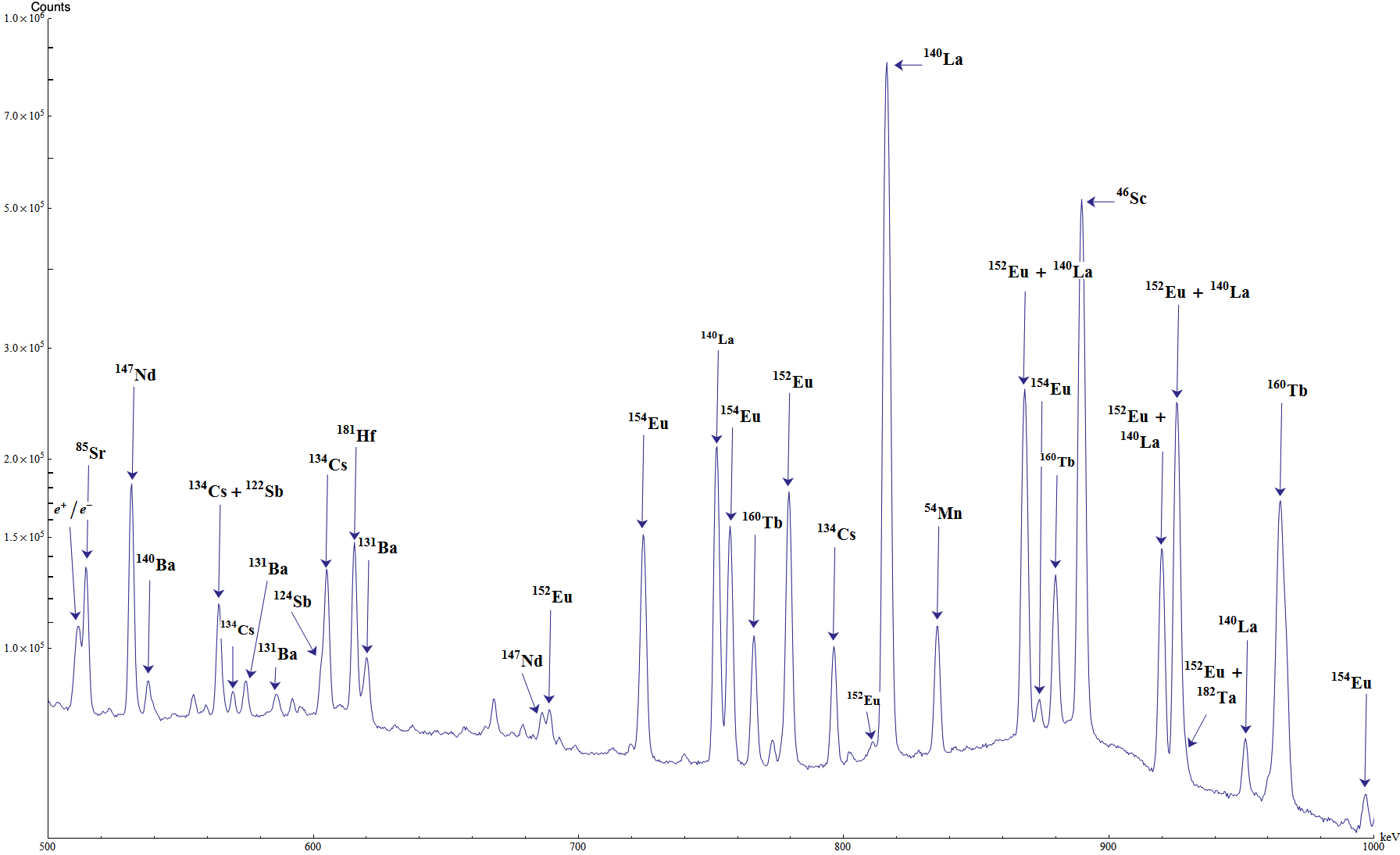}
    \caption[]{Full spectrum of one of the Kilimanjaro samples with most peaks identified.
    }
    \label{espectro Kili1_2 500}
\end{sidewaysfigure}

\begin{sidewaysfigure}[h] 
    \centering
    \includegraphics[width=21cm]{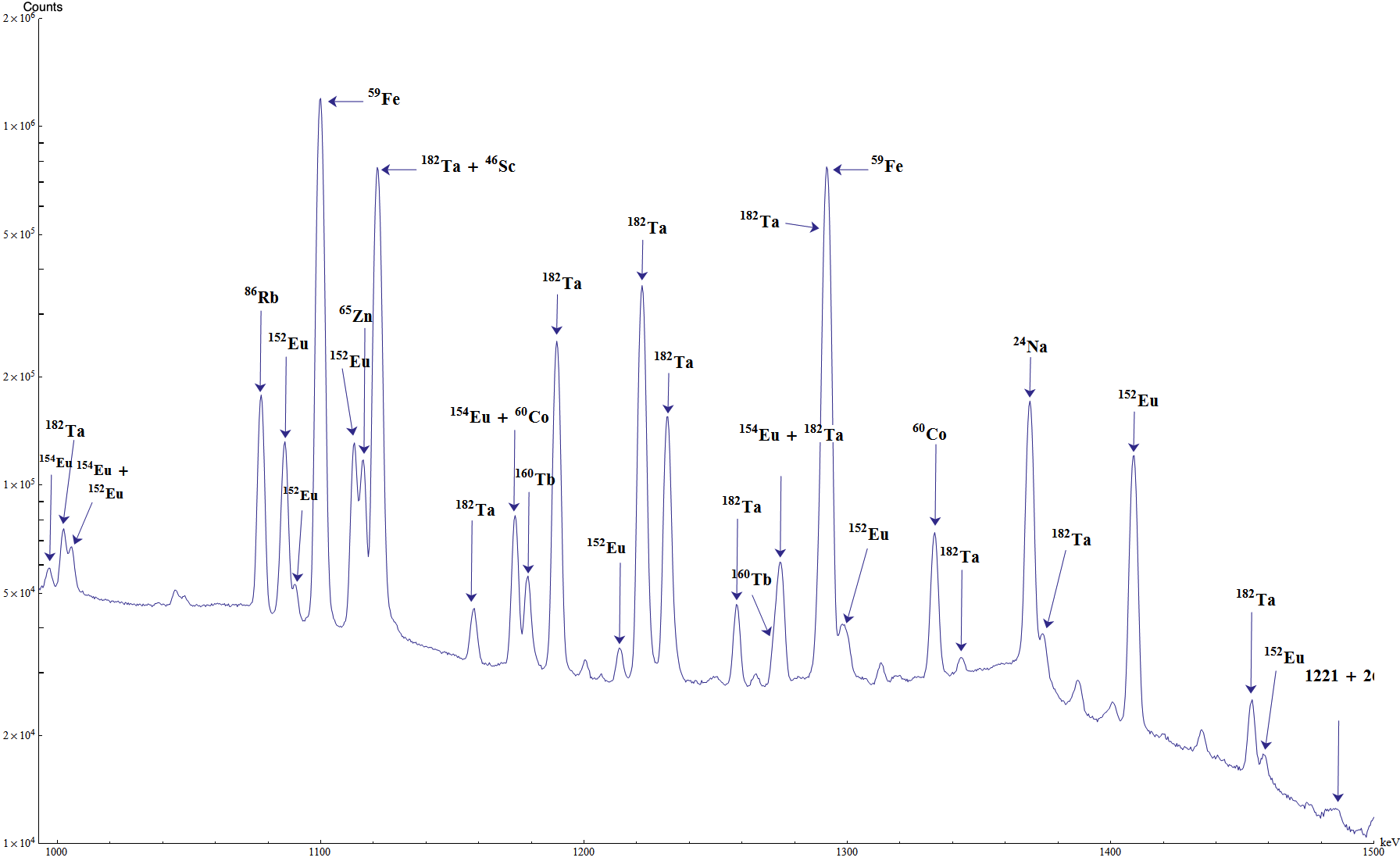}
    \caption[]{Full spectrum of one of the Kilimanjaro samples with most peaks identified.
    }
    \label{espectro Kili1_2 1000}
\end{sidewaysfigure}

\begin{sidewaysfigure}[h]
    \centering
    \includegraphics[width=21cm]{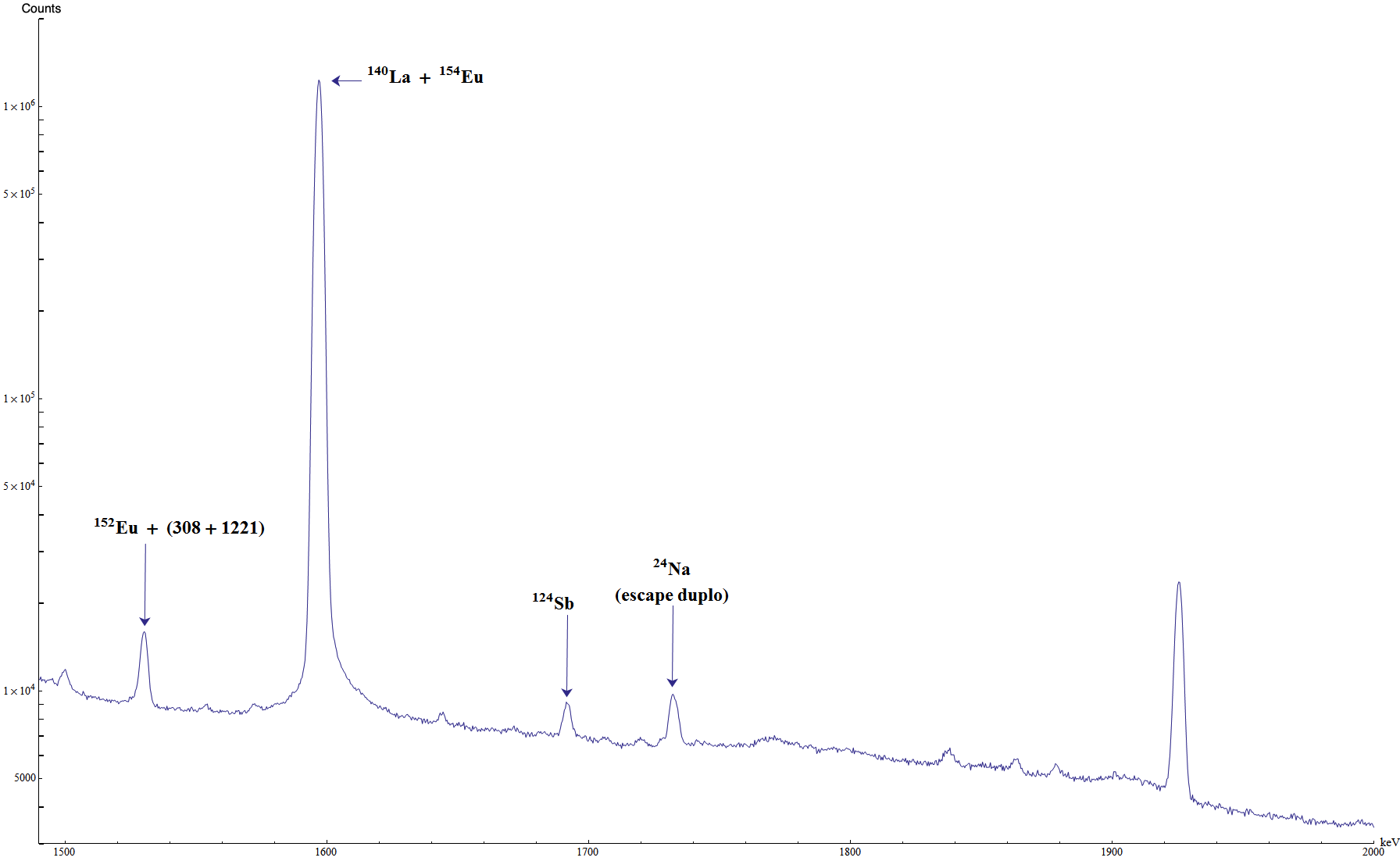}
    \caption[]{Full spectrum of one of the Kilimanjaro samples with most peaks identified.
    }
    \label{espectro Kili1_2 1500}
\end{sidewaysfigure}

\begin{sidewaysfigure} 
    \centering
    \includegraphics[width=21cm]{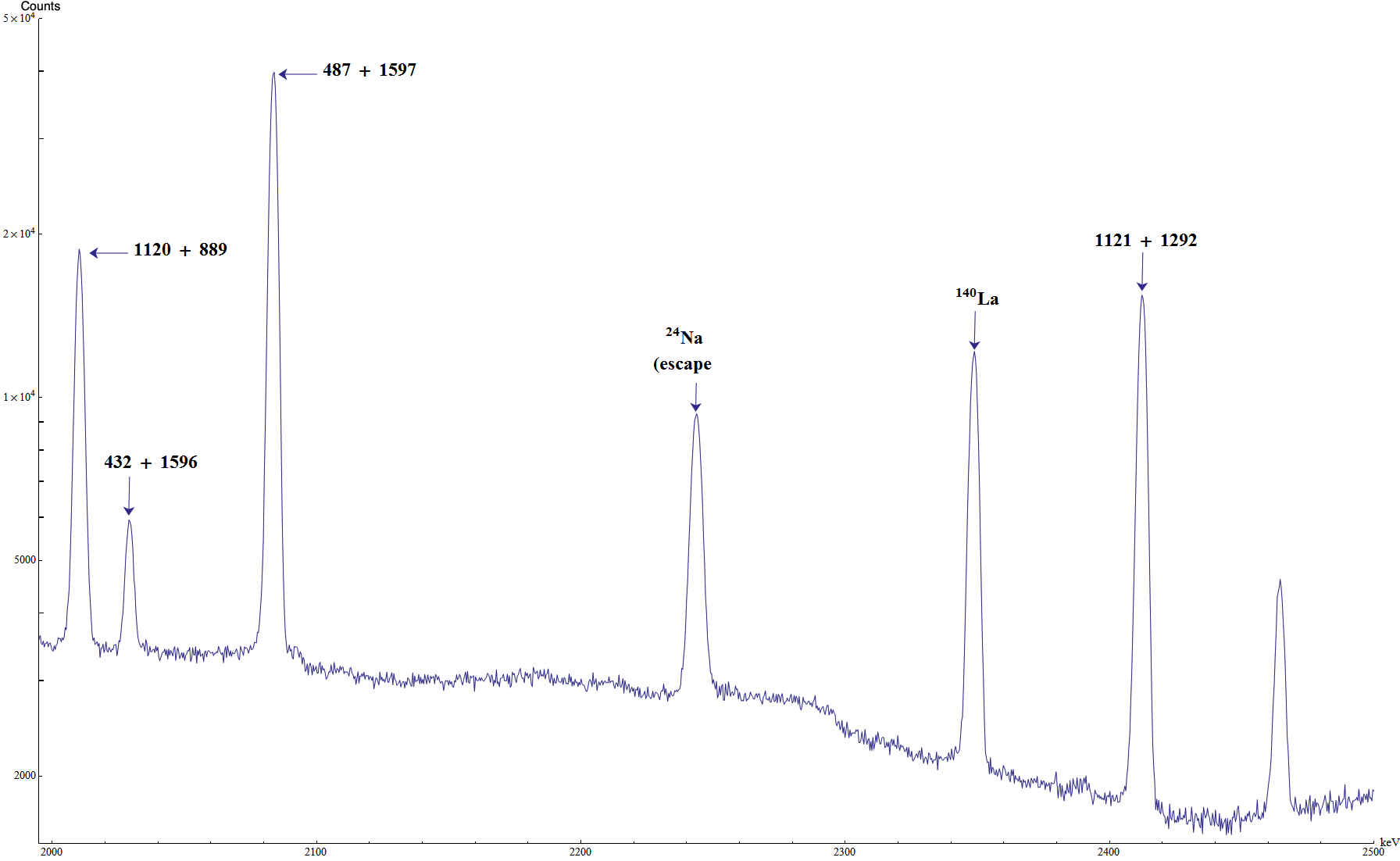}
    \caption[]{Full spectrum of one of the Kilimanjaro samples with most peaks identified.
    }
    \label{espectro Kili1_2 2000}
\end{sidewaysfigure}

\begin{sidewaysfigure}[h] 
    \centering
    \includegraphics[width=21cm]{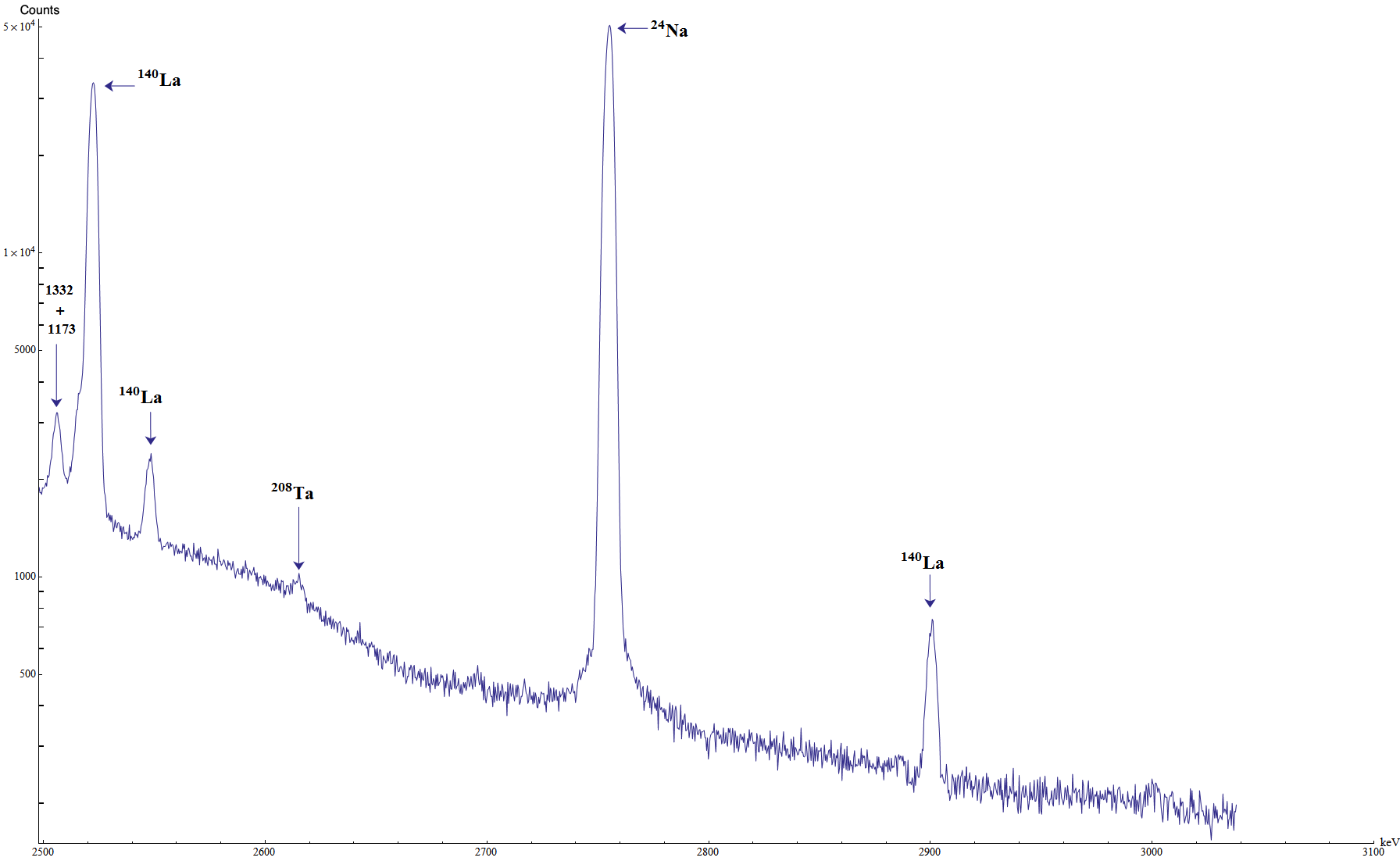}
    \caption[]{Full spectrum of one of the Kilimanjaro samples with most peaks identified.
    }
    \label{espectro Kili1_2 2500}
\end{sidewaysfigure}


\clearpage
\bibliographystyle{elsarticle-harv} 
\bibliography{cas-refs}



\end{document}